# New photonic conservation laws in parametric nonlinear optics

Gavriel Lerner[1,2], Matan Even Tzur[1,2], Ofer Neufeld[1,2,3], Avner Fleischer[4], and Oren Cohen[1,2]

[1]*Physics Department, Technion – Israel Institute of Technology, Haifa, Israel.*
[2]*Solid State Institute, Technion – Israel Institute of Technology, Haifa, Israel.*
[3]*Max Planck Institute for the Structure and Dynamics of Matter, Hamburg, Germany.*
[4]*Chemistry Department, Tel Aviv University, Tel Aviv, Israel.*
Corresponding authors' e-mail addresses: gavriel@campus.technion.ac.il, oren@technion.ac.il

*Abstract*.— Conservation laws are one of the most generic and useful concepts in physics. In nonlinear optical parametric processes, conservation of photonic energy, momenta and parity often lead to selection rules, restricting the allowed polarization and frequencies of the emitted radiation. Here we present a new scheme to derive conservation laws in optical parametric processes in which many photons are annihilated and a single new photon is emitted. We then utilize it to derive two new such conservation laws. Conservation of *reflection-parity* (RP) arises from a generalized reflection symmetry of the polarization in a superspace, analogous to the superspace employed in the study of quasicrystals. Conservation of *space-time-parity* (STP) similarly arises from space-time reversal symmetry in superspace. We explore these new conservation laws numerically in the context of high harmonic generation and outline experimental set-ups where they can be tested.

In *parametric* nonlinear optics (e.g., harmonic generation and parametric amplification), the medium returns to its initial state at the end of the interaction, hence the total radiation energy, momentum and parity are conserved [1]. This leads to so-called photonic conservation laws that are very powerful for analyzing processes in nonlinear optics, especially deriving selection rules that determine which photonic channels are allowed/forbidden, where the allowed channels satisfy all the photonic conservation laws simultaneously [2–8]. (note that the term photon is used, even though these conservation laws do not originate from the quantum nature of light, but from discrete symmetries. That is, they can be derived within a classical theory, as shall be done below) For example, in harmonic generation, energy conservation coerces the energy of a generated photon to equate exactly to the total energy of the annihilated photons. Thus, if the pump consists of only photons at angular frequency ω then emission of non-integer harmonics is forbidden. In another example, the simultaneous conservation of energy, parity and spin angular momentum leads to the selection rule of high harmonic generation driven by co-propagating, bi-chromatic ω-2ω pump fields that are circularly-polarized with opposite helicity [7,9,10].

Another approach for deriving selection rules in nonlinear optics is by analyzing the static and dynamical symmetries (DSs) of the light-matter system. In the field of nonlinear optics, symmetries are standardly used to determine whether a particular nonlinear process is allowed or forbidden according to the medium's point-group [1,11]. Recently, a more general group theory was developed describing the symmetries of the EM field's time-dependent polarization [12], and its interaction with matter [10]. Recent studies utilized DSs to predict many new selection rules, both due to microscopic symmetries [13,14], microscopic-macroscopic symmetries [15–19], and symmetries in synthetic dimensions [20,21]. Such DSs and their associated selection rules have been applied to shaping the waveforms of EUV and X-ray radiation emitted from HHG [8,22–24] and have enabled ultrafast symmetry-breaking spectroscopy of molecular [25,26] and solid orientation [27], molecular symmetries [26], chirality [28–30], imaging of microscopic electric field distributions [31], and detection of topological phase transitions [32]. While the photonic conservation laws approach for deriving selection rules is often more intuitive, the symmetry approach is more general, i.e., it leads to selection rules that cannot be derived by photonic conservation laws. Such examples include reflection DSs that lead to linearly polarized only harmonics, and elliptical DSs that leads to conservation of the polarization ellipticity [10]. We are motivated to bridge this gap by deriving new photonic conservation laws.

Here we present a new method to derive photonic conservation laws, associated with DSs of the pump, in parametric nonlinear optical processes (this derivation cannot be based on Noether's theorem as Floquet DSs are discrete [10]). Our approach is based on superspace representation concept (a representation standardly used in the context of quasicrystals [33], see section 5 of the SI) and the recent multiscale dynamical symmetries concept [15]. Then, we employ this methodology to derive two new conservation laws: reflection parity (RP) and space-time parity (STP). These two conservation laws predict the direction and phase, respectively, of emitted linearly polarized harmonics according to the parity of the harmonic generation process. Finally, we explore these conservation laws numerically in high harmonic generation.

We begin by considering a general nonlinear optical process in which the nonlinear medium is isotropic and stationary and its initial and final states are identical. The driving field consist of $N$ photon-types $\gamma_n$ where $n = 1, \ldots, N$, and each photon type corresponds to a particular frequency and polarization. The type of the emitted photon is denoted by $\gamma_f$ (where $\gamma_f$ is different from all $\gamma_n$). Within this picture the nonlinear process may be represented by the following photonic reaction equation:

$$\sum_n q_n \gamma_n \to \gamma_f \qquad (1)$$

in which $q_n$ is the number of driver photons of type $\gamma_n$ that are annihilated (or generated when $q_n$ is negative) in the generation of a photon of type $\gamma_f$.

The known photonic conservation laws of these processes are given in rows 1-5 in table 1. Rows 6 and 7 present new conservation laws. All seven conservation laws will be derived below or in the SI.

|   | Quantity | Photonic conservation law | Constraints |
|---|---|---|---|
| 1 | energy | $\omega_f = \Sigma_n q_n \omega_n$ | $q_n \in \mathbb{Z}$ |
| 2 | linear momentum | $\vec{k}_f = \Sigma_n q_n \vec{k}_n$ | $|\vec{k}| = n_\omega \omega / c$ |
| 3 | orbital angular momenta | $l_f = \Sigma_n q_n l_n$ | $l \in \mathbb{Z}$ |
| 4 | spin angular momenta | $s_f = \Sigma_n q_n s_n$ | $s = \pm 1$ |
| 5 | parity | $p_f = \Pi_n p_n^{q_n}$ | $p = -1$ |
| 6 | reflection parity | $r_f = \Pi_n r_n^{q_n}$ | $r = \pm 1$ |
| 7 | space-time parity | $u_f = \Pi_n u_n^{q_n}$ | $u = \pm 1$ |

**Table 1.** Photonic conservation laws in the generation process, $\sum_n q_n \gamma_n \to \gamma_f$. Rows 6 and 7 present new conservation laws that are derived below.

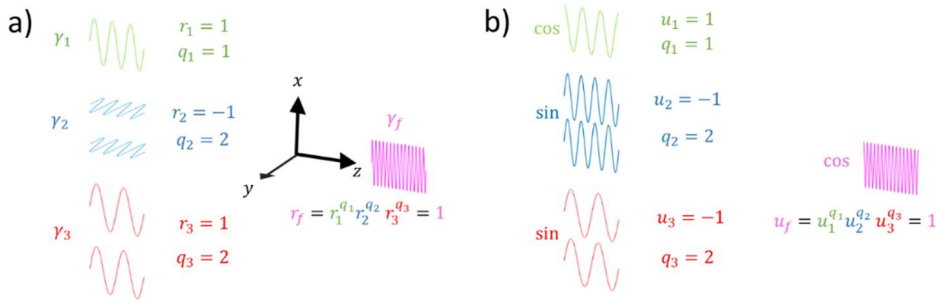

**Figure 1**. Schematic illustrations of (a) conservation of reflection parity that is shown in row 6 in table 1 and (b) space-time parity that is shown in row 7 in table 1.

The constraints on $\omega_n, \vec{k}_n, l_n$, and $s_n$ are given in the right column of Table 1. This set of equations leads to the selection rules of nonlinear optics processes. The constraint on the spin, $s = \pm 1$, leads to selection rules on the polarization of the emitted photons [16,34–36], and the constraint of phase matching is important for interaction regions thicker than the coherence length. We shall now present a method to derive the photonic conservation laws, starting with the known ones and then two novel laws. The method is based on the superspace representation, which was developed in the context of quasicrystals [5]. The first step is to represent the driving field to superspace of N dimensions (as the number of $N$ photon-types $\gamma_n$). Now, in the provided high enough dimensionality, a DS is guaranteed. Using this DSs, we derive the selection rule. This selection rule is general since it is applicable for any driving field. We then formulate the general selection as an intuitive photonic conservation law.

We derive the selection rules from the symmetries according to [15] which reported a theory for multi-scale dynamical symmetries and their selection rules. These multi-scale symmetries consist of operations in time and in both microscopic and macroscopic space scales. Such a symmetry of the electric field, $\vec{E}$, can be described by:

$$\vec{E}(\vec{X}) = \hat{\gamma}\vec{E}(\hat{\Gamma}\vec{X} + \vec{a}) \tag{2}$$

where $\vec{X}$ is the space-time vector, $\vec{a}$ is the macro-space time translation vector. $\vec{X}$ and $\vec{a}$ can be vectors in the physical space or in the superspace (see section 5 of SI). $\hat{\gamma}$ is a microscopic operation, and $\hat{\Gamma}$ is a point-group operation in macro-space time. As shown in [15], harmonic generation in an isotropic medium with a pump exhibiting the above multi-scale dynamical symmetry, exhibits the following selection rule in Fourier domain (if $\vec{a}$ includes translation along the propagation axis then re-absorption of the harmonics is neglected):

$$\hat{\gamma}\vec{F}(\hat{\Gamma}\vec{k}) \exp(i\vec{k} \cdot \vec{a}) = \vec{F}(\vec{k}) \tag{3}$$

where $\vec{k}$ is the space-time wavevector and $\vec{F}(\vec{k})$ is the Fourier coefficient of the generated field, which is a complex-valued vector. In case of a discrete symmetry, eq. (3) leads to:

$$\phi_i(\vec{k}) = \vec{k} \cdot \vec{a} + \alpha_i + \phi_i(\hat{\Gamma}\vec{k}) - 2\pi Q \tag{4}$$

where $\exp(i\alpha_i)$ is the $i$th eigenvalue of the microscopic operation, $Q$ is an integer and $\phi_i(\vec{k})$ is the phase of the Fourier coefficient $\vec{F}(\vec{k})$. In case of a continuous symmetry that involve microscopic rotation of $\delta\alpha$, macroscopic rotation of $\delta 2\pi lm/n$, and $\delta\vec{a}$ translation for any real number $\delta$, eq. (3) leads to:

$$\vec{k} \cdot \vec{a} \pm \alpha + 2\pi lm/n = 0, \tag{5}$$

where $l$ is the allowed winding number that characterizes the orbital angular momentum (OAM) of the emitted $\vec{k}$ harmonic. Next, we apply the proposed method for deriving the selection rules in rows 1-3 in Table 1. Consider a general field composed of superposition of monochromatic plane waves:

$$\vec{E}(t, X_1, X_2, \ldots, X_M) = \sum_{n=1}^{N} \vec{E}_n = \sum_{n=1}^{N} \vec{a}_n e^{i(\omega_n t + \sum_{m=1}^{M} k_{n,m} \cdot X_m)} \tag{6}$$

where $X_m$ are $M$ different orthogonal space dimensions (physically, it is always the case that $M = 3$, however, $M$ is effectively smaller in most experimental realizations, e.g., for plane wave $M = 1$), $\omega_n$ and $k_{n,m}$ are the angular frequency and wavevector of $\vec{E}_n$, respectively. For transforming into a superspace representation, we add more dimensions such that the field will be periodic for any combinations of $\omega_n$ and $k_{n,m}$. The field in superspace is given by:

$$\vec{E}(t_1, t_2, \ldots, t_N, X_{1,1}, X_{1,2}, \ldots, X_{2,1}, X_{2,2}, \ldots, X_{N,M}) = \sum_{n=1}^{N} \vec{a}_n e^{i(\omega_n t_n + \sum_{m=1}^{M} k_{n,m} \cdot X_{n,m})} \tag{7}$$

This field is periodic with respect to $t_n$ with periodicity $2\pi/\omega_n$. Inserting the symmetry $t_n \to t_n + 2\pi/\omega_n$ to eq.(4) with the translation vector $\vec{a} = \frac{2\pi}{\omega_n}\hat{t}_n$ where $\hat{t}_n$ is a unit vector along the $t_n$ axis and using the superspace wave-vector $\vec{k}_S = \sum_{n=1}^{N} \omega_n^{(g)} \hat{t}_n + \sum_{m=1}^{M} \sum_{n=1}^{N} k_{n,m}^{(g)} \hat{X}_{n,m}$ where $\omega_n^{(g)}$ and $k_{n,m}^{(g)}$ are the generated temporal and spatial frequencies in superspace, gives:

$$\phi_i(\vec{k}) = \vec{k}_S \cdot \vec{a} + \alpha_i + \phi_i(\hat{\Gamma}\vec{k}) - 2\pi Q = \omega_n^{(g)} \cdot \frac{2\pi}{\omega_n} + \phi_i(\vec{k}) - 2\pi Q$$
$$\rightarrow \omega_n^{(g)} = Q\omega_n = q_n \omega_n \tag{8}$$

Thus the allowed frequencies along the $\hat{t}_n$ axis are harmonics of $\omega_n$ with $q_n$ possible harmonics order. Hence, the allowed superspace temporal frequencies are any combination of $q_n$ integers $\vec{\omega}_{\vec{q}} = \sum_{n=1}^{N} q_n \omega_n \hat{t}_n$. Projecting $\vec{\omega}_{\vec{q}}$ back to physical space ($t_n = t$), the allowed temporal frequencies are $\omega_{\vec{q}} = \sum_{n=1}^{N} q_n \omega_n$. Similar symmetry appears for each spatial coordinates, $X_m$, which lead to spatial harmonics, $k_{m\vec{q}_m} = \sum_{n=1}^{N} q_{n,m} k_{n,m}$. Next, we will show that $\vec{q}_m = \vec{q}$ by a symmetry that connects the spatial harmonics to the temporal harmonics. This is a continuous symmetry of $t_n \rightarrow t_n + \frac{\delta}{\omega_n}$ and $X_{n,m} \rightarrow X_{n,m} - \frac{\delta}{k_{n,m}}$ (i.e. $\vec{k} \cdot \vec{a} = \frac{\omega_n^{(g)}}{\omega_n} - \frac{k_{n,m}^{(g)}}{k_{n,m}}$), which according to eq. (5)

$$\vec{k} \cdot \vec{a} = \frac{\omega_n^{(g)}}{\omega_n} - \frac{k_{n,m}^{(g)}}{k_{n,m}} = \frac{q_n \omega_n}{\omega_n} - \frac{q_{n,m} k_{n,m}}{k_{n,m}} = 0$$

gives the selection rule, $q_n - q_{n,m} = 0 \rightarrow q_n = q_{n,m}$. This important selection rule implies that the numbers of annihilated photons of each driver, with respect to energy and linear momentum (when $X_m$ is in Cartesian coordinates) or orbital angular momentum (when $X_m$ is the cylindrical angle in cylindrical coordinates), conservation rules are the same. Hence, we obtained rows 1-3 in table 1. Therefore the allowed superspace wave-vector of the emitted field is of the form:

$$\vec{k}_S = \sum_{n=1}^{N} q_n \omega_n \hat{t}_n + \sum_{m=1}^{M} \sum_{n=1}^{N} q_n k_{n,m} \hat{X}_{n,m}. \tag{9}$$

The spin angular momentum (parity) conservation law in row 4 (5) in table 1 is derived in section 1 (3) of the SI.

Next, we derive the new conservation law that is shown in row 6 of table 1 (see also schematic illustration in Fig. 1a). Consider a general EM field with polarization in the x-y plane, written here as a superposition of waves with x or y linear polarization:

$$\vec{E}(t, X_1, X_2, \dots) = \hat{x} \sum_{n_x=1}^{N_x} a_{n_x} e^{i(\omega_{n_x} t + \sum_{m=1}^{M} k_{n_x,m} \cdot X_m)} + \hat{y} \sum_{n_y=1}^{N_y} a_{n_y} e^{i(\omega_{n_y} t + \sum_{m=1}^{M} k_{n_y,m} \cdot X_m)} \tag{10}$$

This can be written in superspace as

$$\vec{E}\left(t_{1_x}, t_{2_x}, \dots, t_{N_x}, t_{1_y}, t_{2_y}, \dots, t_{N_y}, X_1, X_2, \dots, X_M\right)$$
$$= \hat{x} \sum_{n_x=1}^{N_x} a_{n_x} e^{i(\omega_{n_x} t_{n_x} + \sum_{m=1}^{M} k_{n_x,m} \cdot X_m)} + \hat{y} \sum_{n_y=1}^{N_y} a_{n_y} e^{i(\omega_{n_y} t_{n_y} + \sum_{m=1}^{M} k_{n_y,m} \cdot X_m)} \tag{11}$$

This field has two discrete symmetries: (I) $t_{n_x} \rightarrow t_{n_x} + \frac{\pi}{\omega_{n_x}}$ for all $n_x = 1 \dots N_x$ simultaneously, $\hat{\sigma}_x$. (II) $t_{n_y} \rightarrow t_{n_y} + \frac{\pi}{\omega_{n_y}}$ for all $n_y = 1 \dots N_y$ simultaneously, $\hat{\sigma}_y$.

Therefore, according to eq. (3) with $\vec{a} = \sum_{n_x=1}^{N_x} \frac{\pi}{\omega_{n_x}} \hat{t}_{n_x}$ for symmetry (I) and $\vec{a} = \sum_{n_y=1}^{N_y} \frac{\pi}{\omega_{n_y}} \hat{t}_{n_y}$ for symmetry (I) and $\vec{k}_S$ of eq. (9), the Fourier domain of the generated field in the wave-mixing process must obey the following equations:

$$\hat{\sigma}_x \vec{F}(\vec{k}) \exp\left(i\pi \sum_{n_x=1}^{N_x} q_{n_x}\right) = \vec{F}(\vec{k})$$
$$\hat{\sigma}_y \vec{F}(\vec{k}) \exp\left(i\pi \sum_{n_y=1}^{N_y} q_{n_y}\right) = \vec{F}(\vec{k})$$
(12)

which dictates that harmonics with odd $\sum_{n_x=1}^{N_x} q_{n_x}$ and even $\sum_{n_y=1}^{N_y} q_{n_y}$ are x-polarized, harmonics with even $\sum_{n_x=1}^{N_x} q_{n_x}$ and odd $\sum_{n_y=1}^{N_y} q_{n_y}$ are y-polarized, and all other harmonics are forbidden. This result is a new photonic conservation law, which is an extension of parity conservation. We can associate this conserved quantity with the following conservation law: each photon carries a RP of $r = +1$ for an x-polarized photon or $r = -1$ for a y-polarized photon. The RP (i.e., the x- or y-polarization) of the emitted photon is $r_f = \prod_n r_n^{q_n}$ which corresponds to the conservation of RP in the sixth row of table 1. Hence, all generated harmonic channels will be linearly polarized and will have phase differences between each other.

Last, we derive the conservation law in row 7 of table 1 (see also schematic illustration in Fig. 1b). Consider a general EM field, written here as superposition of cosine and sine waves with a linear polarization:

$$\vec{E}(t, X_1, X_2, \dots) = \sum_{n_c=1}^{N_c} \vec{a}_{n_c} \cos\left(\omega_{n_c} t - \sum_{m=1}^{M} k_{n_c,m} \cdot X_m\right)$$
$$+ \sum_{n_s=1}^{N_s} \vec{a}_{n_s} \sin\left(\omega_{n_s} t - \sum_{m=1}^{M} k_{n_s,m} \cdot X_m\right)$$
(13)

where $\vec{a}_{n_c}$ and $\vec{a}_{n_s}$ are linearly polarized real-valued amplitudes. In superspace, we can write the field as:

$$\vec{E}(t_{1_c}, t_{2_c}, \dots, t_{N_c}, t_{1_s}, t_{2_s}, \dots, t_{N_s}, X_1, X_2, \dots, X_M)$$
$$= \sum_{n_c=1}^{N_c} \vec{a}_{n_c} \cos\left(\omega_{n_c} t_{n_c} - \sum_{m=1}^{M} k_{n_c,m} \cdot X_m\right)$$
$$+ \sum_{n_s=1}^{N_s} \vec{a}_{n_s} \sin\left(\omega_{n_s} t_{n_s} - \sum_{m=1}^{M} k_{n_s,m} \cdot X_m\right)$$
(14)

This field has a discrete symmetry, $t_{n_s} \to t_{n_s} + \frac{\pi}{\omega_{n_s}}$ for all $n_s = 1 \dots N_s$ simultaneously, $\hat{I}$, where $\hat{I}$ is the STP operator (i.e., $\hat{I}\vec{E}(\vec{X}) = \vec{E}(-\vec{X})$). Hence, $\vec{a} = \sum_{n_s=1}^{N_s} \frac{\pi}{\omega_{n_s}} \hat{t}_{n_s}$ and $\hat{I}\vec{k} = \hat{I}\vec{k} = -\vec{k}$. Thus inserting eq. (4), the symmetry $t_{n_s} \to t_{n_s} + \frac{\pi}{\omega_{n_s}}$, $\hat{I}$ gives:

$$\phi_i(\vec{k}) = \vec{k}_S \cdot \vec{a} + \alpha_i + \phi_i(\hat{I}\vec{k}) - 2\pi Q = \pi \sum_{n_s} q_{n_s} + \phi_i(-\vec{k}) - 2\pi Q$$
$$\to \pi \sum_{n_s} q_{n_s} + \phi_i(-\vec{k}) - \phi_i(\vec{k}) = 2\pi Q$$
(15)

where $\phi_i(-\vec{k}) - \phi_i(\vec{k}) = 0$ for a generated cosine mode, and $\phi_i(-\vec{k}) - \phi_i(\vec{k}) = \pi$ for a generated sine mode. Therefore, for a generated sine mode in the induced polarization, $\sum_{n_s} q_{n_s}$ must be odd, and for a cosine mode, $\sum_{n_s} q_{n_s}$ must be even.

This result is a new photonic conservation law, which is another extension to parity conservation. We can describe this law by considering photons that carry a STP, $u$, that equals 1 for cosine modes or -1 for sine modes. This STP is conserved such that the $u_f$ of the emitted photon is: $u_f = \prod_n u_n^{q_n}$, which corresponds to the conservation of the STP in the seventh row of table 1. Hence, all generated harmonic channels will be linearly polarized and will have phase differences between each other.

*Numerical investigations.* — Below, we present numerical investigations of the selection rules that result from the new photonic conservation laws. These selection rules cannot result from the photonic conservation laws in rows 1-5 in table 1. The single-atom HHG spectra are calculated using the strong field approximation method [37] in two spatial dimensions, for an atom with a hydrogen-like dipole [38] and ionization potential of Argon (15.76 eV), with the use of saddle point approximation for momentum integrations and numerical integration in the ionization time domain.

*Reflection parity conservation law (row 6 in table 1).* — Consider the driving field:

$$\vec{E}(t) = \sqrt{I_0}A(t)[\hat{x}\cos(\omega_1 t) + \hat{y}\cos(\omega_2 t + \phi)] \qquad (16)$$

The generated frequencies are $\omega_f = q_1\omega_1 + q_2\omega_2$. According to the RP conservation law, annihilation of odd $q_1$ and even $q_2$ photons leads to the emission of an $\hat{x}$ polarized photon, while annihilation of even $q_1$ and odd $q_2$ photons leads to the emission of a $\hat{y}$ polarized photon. We test this prediction numerally with $\omega_1 = 2.35 \cdot 10^{15}\ rad/s$ (λ=800nm) and $\omega_2 = \sqrt{2}\omega_1$ (and $\phi = 1\ rad$), which allows us to easily identify the channel $(q_1, q_2)$ corresponding to each spectral line in the emitted power spectrum. The amplitude envelope is $A(t) = \exp\left(-\left(\frac{\omega_1}{2\pi \cdot 30}t\right)^6\right)$, and the peak intensity is $I_0 = 10^{14}\ W/cm^2$.

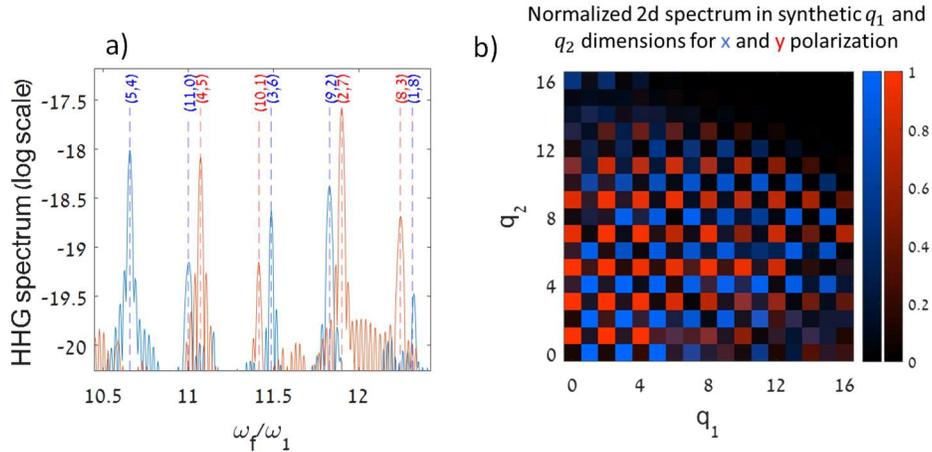

**Figure 2.** a) The power spectrum of x (blue) and y (red) linearly polarized harmonics driven by the field in eq. (16), demonstrating the RP photonic law (row 6 in table 1) The polarization of the channels $(q_1, q_2)$ predicted by the conservation law is displayed by the dashed lines. b) The normalized power spectrum of x (blue) and y (red) linearly polarized harmonics driven by the field in eq. (16) as a function of $q_1$ and $q_2$. As shown, the numerical results correspond well to the prediction.

As shown in fig. 2, the polarizations of the numerically generated high harmonics agree with the prediction of the conservation law.

*Space-time parity (row 7 in table 1).* — Consider the driving field:

$$\vec{E}(t) = \sqrt{I_0}A(t)(\hat{x}\cos(\omega_1 t) + \hat{y}\sin(\omega_2 t) + (0.5\hat{x} + \hat{y})\cos(\omega_3 t)) \qquad (17)$$

The generated frequencies are $\omega_f = q_1\omega_1 + q_2\omega_2 + q_3\omega_3$. According to the STP photonic law, when $q_1 + q_3$ is odd and $q_2$ is even, the harmonic field is proportional to $\cos(\omega_f t)$, while it is proportional to $\sin(\omega_f t)$ when $q_1 + q_3$ is even and $q_2$ is odd. We test this prediction numerally with $\omega_1 = 2.35 \cdot 10^{15}\ rad/s$ ($\lambda$=800nm), $\omega_2 = 2\omega_1$ and $\omega_3 = 3\omega_1$. The peak intensity is $I_0 = 10^{12}\ W/cm^2$. In this case, all of the odd harmonics (of $\omega_1$) are cosine functions, while all of the even harmonics are sine functions. We simulate the induced dipole as a function of time and then calculate the cosine (sine) harmonics by taking the real (imaginary) part of the Fourier transform of the induced dipole for both (arbitrary) orthogonal linear polarizations. The results are presented in fig. 3a), showing that harmonics in the plateau region indeed correspond to the predicted selection rule, up to deviations that are smaller than 0.1%.

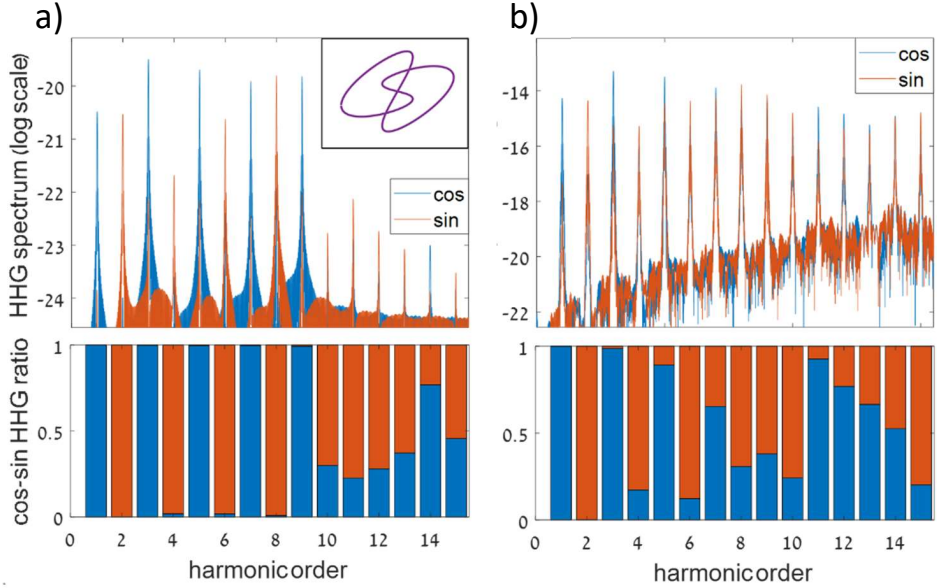

**Figure 3.** Numerical investigation of the STP photonic law. a) The lissajous curve of the pump is plotted in the top right inset and is given in eq. (17) with intensity, $I_0 = 10^{12}\ W/cm^2$. The figure shows the power spectrum of high harmonics with cosine (blue) and sine (red) functions. The bottom plot shows the normalized relative power to the sine and cosine fields of each harmonic. As predicted by the conservation law, odd harmonics are cosine functions while even harmonics are sine. Harmonics beyond the cutoff exhibit significant deviations from the predicted conservation law. b) Numerical investigation of the STP photonic law at relatively large ionization: The pump is the same as in a), but with higher intensity, $I_0 = 10^{14}\ W/cm^2$. The figure shows the power spectrum of high harmonics with cosine (blue) and sine (red) functions, showing that the STP is broken even for harmonics in the plateau.

Notably, time reversal symmetry is broken by ionization process; hence, the STP photonic law, which results from time reversal symmetry, is sensitive to ionization and is not perfectly upheld. To test this assumption, we increased the intensity of the driver field to $I_0 = 10^{14}\ W/cm^2$. Now, as shown in Fig. 3b, only harmonics 1-6 behave according to the conservation law, while the higher harmonics do not.

***Discussion.*** — To conclude, we explored here photonic conservation laws arising from symmetries in a superspace representation of the electromagnetic field interacting with matter, including two new conservation laws. These two laws describe the conservation of polarization-like properties of light, as well as its phase-like properties, and complement the well-established conservation of spin and orbital angular momentum law. Beyond the development of the methodology and the derivation of new photonic conservation laws, we highlight that our work assigns new photonic characters associated with the conservation laws to electromagnetic waves, which should motivate future research to identify their origin and connection to other photonic characteristics.